# THE IMPACT OF AI TOOL ON ENGINEERING AT ANZ BANK
# AN EMPIRICAL STUDY ON GITHUB COPILOT WITHIN CORPORATE ENVIRONMENT

Sayan Chatterjee[1], Ching Louis Liu[2], Gareth Rowland, Tim Hogarth

[1]The Australia and New Zealand Banking Group Limited, Melbourne, Australia
sayan.chatterjee@anz.com
[2]The Australia and New Zealand Banking Group Limited, Melbourne, Australia
louis.liu@anz.com

## ABSTRACT

*The increasing popularity of AI, particularly Large Language Models (LLMs), has significantly impacted various domains, including Software Engineering. This study explores the integration of AI tools in software engineering practices within a large organization. We focus on ANZ Bank, which employs over 5000 engineers covering all aspects of the software development life cycle. This paper details an experiment conducted using GitHub Copilot, a notable AI tool, within a controlled environment to evaluate its effectiveness in real-world engineering tasks. Additionally, this paper shares initial findings on the productivity improvements observed after GitHub Copilot was adopted on a large scale, with about 1000 engineers using it.*

*ANZ Bank's six-week experiment with GitHub Copilot included two weeks of preparation and four weeks of active testing. The study evaluated participant sentiment and the tool's impact on productivity, code quality, and security. Initially, participants used GitHub Copilot for proposed use-cases, with their feedback gathered through regular surveys. In the second phase, they were divided into Control and Copilot groups, each tackling the same Python challenges, and their experiences were again surveyed. Results showed a notable boost in productivity and code quality with GitHub Copilot, though its impact on code security remained inconclusive. Participant responses were overall positive, confirming GitHub Copilot's effectiveness in large-scale software engineering environments. Early data from 1000 engineers also indicated a significant increase in productivity and job satisfaction.*

## 1. INTRODUCTION

Generative AI unleashes the next wave of productivity through operational efficiency and quicker-informed decisions. In the field of Software Engineering, research suggests that developers use generative Artificial Intelligence (AI) as a pair-programmer to increase the output of high-quality code. By augmenting employees' capacity, more time and resources can be allocated towards innovation. AI, however, raises inherent risks, uncertainties and unintentional consequences regarding intellectual property, data security and privacy. It is, therefore, crucial to measure the quantitative and qualitative benefits of AI Tools prior to large scale adoption.

Despite the presence of various AI-assisted tools in the marketplace, such as Code Whisperer [1]. GitHub Copilot stands out as one of the pioneers in this domain. This early entry into the market, coupled with its robust features, has informed our decision to investigate in GitHub Copilot.

GitHub Copilot functions as an advanced assistant for software developers, powered by artificial intelligence (AI). It is adept at generating syntactically correct and contextually relevant code snippets across a diverse array of programming languages such as Java, Python, C#, C++, and others [7]. The tool not only produces code but is also capable of generating comprehensive comments that elucidate the purpose and functionality of the code, given the current context within the development environment.

GitHub Copilot is compatibility with a range of Integrated Development Environments (IDEs), including Visual Studio Code, Neovim, JetBrains suite, and GitHub Codespaces [3]. Initially launched in a technical preview on Visual Studio Code on June 29, 2021, GitHub offer Copilot as a subscription service on June 21, 2022, accessible to both individual developers and corporate entities.

The underlying architecture of Copilot is predicated on Generative Pre-trained Transformer (GPT) technology, which has been meticulously refined using publicly available GitHub code to enhance its code recognition and generation capabilities.

The GitHub Copilot experiment is an Architecture & Engineering initiative with two main objectives. Firstly, to establish this experiment as an example of a purposeful and methodical guide for the adoption of AI pair programming technologies ranging from experimental to large-scale at ANZ Bank. The second objective is to collect statistical measures on engineers' productivity, code quality, and the security of the code generated while using GitHub Copilot.
This paper offers the following contributions:

- A systematic examination of productivity improvements regarding code quality, development time, and problem complexity within a corporate environment.
- An analysis of engineer sentiment towards the use of AI-aided tools in the software development process.
- An initial validation of GitHub Copilot's efficacy post-production, informed by data from approximately 1000 engineers.

The remainder of this paper is structured as follows: Section 2 provides a succinct review of the related literature. Section 3 details the study's design and methodology. The data collection and subsequent analysis are presented in Section 4. Finally, Section 5 delves into the discussion of our findings and outlines potential avenues for future research.

## 2. RELATED WORK

Research on GitHub Copilot's application in Software Engineering tasks, such as code generation, testing, security, and documentation, has been expanding. These studies provide valuable insights into the tool's effectiveness and areas of improvement.

A pivotal experiment by Microsoft [2] in 2022 involved 95 engineers working in a realistic environment. They were tasked with creating an HTTP server in JavaScript, with the option to seek online help for challenges faced. Engineers were divided into two groups: one with access to GitHub Copilot (the treatment group) and one without (the control group). The results were significant: the treatment group completed their tasks 55.8% faster, with developers who had less experience, older programmers, or those who programmed more frequently seeing the most benefits. Our ANZ Bank experiment, although involving different tasks and Python as the programming language, aligns with these findings.

In another study, B Yetiştiren et al. [3] compared GitHub Copilot with AWS CodeWhisperer and ChatGPT across various parameters including validity, correctness, security, reliability, and maintainability. They found that the latest versions of these tools varied in their ability to generate correct code, with GitHub Copilot showing an 18% improvement in newer versions. Notably, GitHub Copilot performed better than CodeWhisperer in an engineering-specific context, while ChatGPT was identified as more of a general-purpose tool. Our study at ANZ Bank similarly observed that GitHub Copilot consistently generated code suggestions in a timely manner, regardless of task complexity, contrasting with the varying time human engineers need for more complex tasks.

An earlier study in 2022 [4] also highlighted a positive sentiment among engineers regarding productivity improvements with GitHub Copilot, a finding echoed in our research.

S Imai [6] conducted an intriguing experiment comparing pair programming with Copilot versus a human partner. The study measured productivity by lines of code produced and code quality by lines removed. While Copilot generated the most lines of code, it also had the most lines deleted. It's important to note that lines of code aren't necessarily indicative of code quality.

In another study, Dakhel et al. [5] set out to determine whether GitHub Copilot positively or negatively affects programmer productivity. Their research comprised two distinct programming tasks: (i) evaluating Copilot's ability to generate correct and efficient solutions for fundamental algorithmic problems such as sorting and implementing data structures, and (ii) contrasting Copilot's solutions with those provided by human programmers across a range of programming challenges.

For the first task, Copilot demonstrated a remarkable ability to tackle most fundamental algorithmic problems, though some of its solutions were found to be buggy and not always reproducible. The second task involved a comparative analysis using a dataset of programming problems with existing human solutions. The study revealed that, while humans generally provided more correct solutions than Copilot, the errors in Copilot's outputs were typically less complex and easier to rectify.

Crucially, Dakhel et al. [5] concluded that the utility of GitHub Copilot varies depending on the user's expertise. For experienced developers, Copilot can be a valuable asset, offering suggestions of a quality comparable to human contributions. Conversely, for novice programmers, the tool poses risks due to their potential inability to discern and correct Copilot's less optimal or erroneous solutions.

## 3. STUDY DESIGN

The GitHub Copilot Experiment ran over six weeks from mid-June'23 to the end of July'23: it consisted of two weeks of preparation, followed by four weeks of experiment execution. The experiment examined the sentiment that participants felt towards GitHub Copilot's Visual Studio Code extension as well as the impact the tool had on participants' productivity, code quality, and code security.

Prior to starting the experiment, risks related to intellectual property, data security and privacy were assessed in conjunction with ANZ's legal and security teams to arrive at a set of guidelines. The scope of the experiment and report will discuss the key aims as follows:
- Do the developers at ANZ feel positive and empowered by having access to Copilot?
- How much does access to this tool make employees work faster, if at all?
- Does this tool make developers at ANZ output better?
- Is the code suggested by Copilot secure?

Additionally, the experiment seeks to answer other meaningful questions:
- How often are Copilot's suggestions considered useful and accepted?
- Does the code suggested by Copilot follow best practices?

Detailed considerations are outlined in a Playbook which participants were required to read and agree to before commencing the experiment.

During the designing of the experiment, the following decisions were made, and constraints considered:
- GitHub Copilot gathers a large breadth of metrics when integrated with Visual Studio Code Integrated Development Environment (IDE). To mitigate possible variability, Visual Studio Code was selected as the single IDE for this experiment to be conducted within.
- Based on the widespread usage of Python programming language in the software engineering industry, as well as accessibility to novice programmers, Python was selected as the only language to be used during statistical hypothesis testing in weeks 3 and 4.
- Given the limited time participants had for the experiment, we chose algorithmic questions for the weeks 3 and 4 code problems instead of application development scenarios typical of workplace challenges.
- Participants were not actively monitored during the experiment. When assessing productivity, we relied on self-reported time and feedback about experience, rather than the exact duration spent in participation.

**A/B Testing:**
To statistically analyse the effectiveness of engineers using Copilot compared to engineers who hand code, A/B Testing was performed by dividing participants into Control group (Copilot extension disabled) and Copilot group (Copilot extension enabled) with the following hypotheses:

$H_0$ = *There is no significant difference in productivity or code quality of engineers using Copilot*
$H_1$ = *there is a statistically significant difference in productivity and code quality of engineers using Copilot*

Participants were divided randomly in half based on employees who had submitted the baseline survey in the preparation stage with an initial minimum expectation of 60 participants. The Control and Copilot group in week 3 were reversed in week 4 so groups remained the same but their ability to enable the copilot extension was changed. To note, the Control group were not permitted to use Copilot but other tools currently available to developers such as searching the internet or using Stack Overflow were allowed.

Six algorithmic coding challenges were provided to solve each week, All participants were expected to attempt the same questions each week, 12 questions in total over week 3 and 4. A key constraint was all engineers were asked to use only Python to code, for uniformity in assessing code quality and interpretation of code correctness, ensuring rigor in statistical evaluation.

On completion, participants uploaded their solutions to their repositories, as well as submitted a survey for each challenge to collect the following metrics:

Productivity
- Total time spent solving a problem (minutes, self-reported)
- Quality
- Unit test success ratio

Bugs
- Code smells

Security
- Code vulnerability

A briefing session was organised on the first day of Phase 2 to ensure participants were all in alignment with task requirements and to establish a support channel if required. Emails were sent every two days to remind participants to submit surveys.

## 4. DATA COLLECTION APPROACH

To analyse data on developer productivity, quality of work performed and sentiment in using Copilot, we decided to use four sources to obtain metrics:

### 4.1. GitHub Copilot

GitHub metrics were selected to provide statistical information relating to usage of the tool and how useful Copilot was in terms of predicting code written. The data would be collected at the end of each week of the experiment, and would track:
- The number of times a suggestion was provided.
- The number of times a suggestion was accepted.
- The number of times all suggested lines of code were accepted (partial acceptance not included)
- The number of languages used and their acceptances rates for suggestions.
- Percentage acceptance rate

### 4.2. Surveys

Online surveys were conducted to obtain data on productivity and sentiment.
- The 'Baseline Survey' was to be completed prior to initiation of phase 1, during the preparation stage.
- The 'Week 1-2 Survey' was to be completed every second day during phase 1 of the experiment.
- The 'Week 3 Survey' and 'Week 4 Survey' was to be completed each time a participant completed a challenge question (up to 12 times in total).

### 4.3. Static Code Analysis

Static code analysis was performed using SonarQube to capture metrics related to code quality and security vulnerabilities

### 4.4. Grading of correctness

Solutions submitted by each participant were graded by the ANZ Copilot Experiment Team according to their correctness, to assess how well they completed the assigned tasks.

### 4.5. Demographics

There are over 100 participants involved in the experiment and their roles are mainly:

1. Software Engineers
2. Cloud Engineers
3. Data Engineers

### 4.6. Statistical Inference

From Phase 2 surveys (A/B Testing), total 200 data points were collected and were used for statistical analysis. Data were also collected by running unit test scripts and SonarQube code review against the code submitted by participants to ANZ GitHub repositories.

Table 1. Summary of Control Group data and Copilot Group data

| | | | |
|---|---|---|---|
| Total_Time_Spent (minutes) | Min. | 4 | 2 |
| | 1st Quartile | 15 | 5 |
| | Median | 20 | 10 |
| | Mean | 30.98 | 17.86 |
| | 3rd Quartile | 35 | 20 |
| | Max. | 150 | 150 |
| Python_Proficiency | Beginner | 10 | 9 |
| | Novice | 4 | 3 |
| | Intermediate | 40 | 51 |
| | Advanced | 22 | 22 |
| | Expert | 6 | 5 |
| Debugging_Time_Ratio | 0-20% | 24 | 48 |
| | 21-40% | 26 | 19 |
| | 41-60% | 19 | 13 |
| | 61-80% | 13 | 8 |

|                  | 81-100%   | 0  | 2  |
|------------------|-----------|----|----|
| Difficulty_Level | Very Easy | 11 | 11 |
|                  | Easy      | 52 | 59 |
|                  | Medium    | 15 | 14 |
|                  | Hard      | 4  | 6  |

Below are the observations from Table 1:

- Both Control Group and Copilot Group have almost equal proportion data points.
- Both mean, median, 1st Quartile and 3rd Quartile values of total time taken to solve a problem (Total_Time_Spent) are significantly less for Copilot Group than Control group.
- Number of data points across Python Proficiencies and Difficulty Levels are well distributed across both Control Group and Copilot Group.
- Debug Percentage is less for Copilot Group compared to Control group

### 4.7. Data Summary

From 200 data points, after quick analysis the following data points were removed.
There were 6 duplicate data points. Duplicate data points are those where a participant submitted two surveys for the same coding problem.

There were 22 data points where participants could not solve the problems. Remaining 172 data points were used for all the statistical analysis presented on this document.

The entire dataset was split into two groups - Control Group and Copilot Group. Below is the summary of the dataset for both Control Group and Copilot Group. Please note that the summary table below only includes numeric columns from Phase 2 surveys.

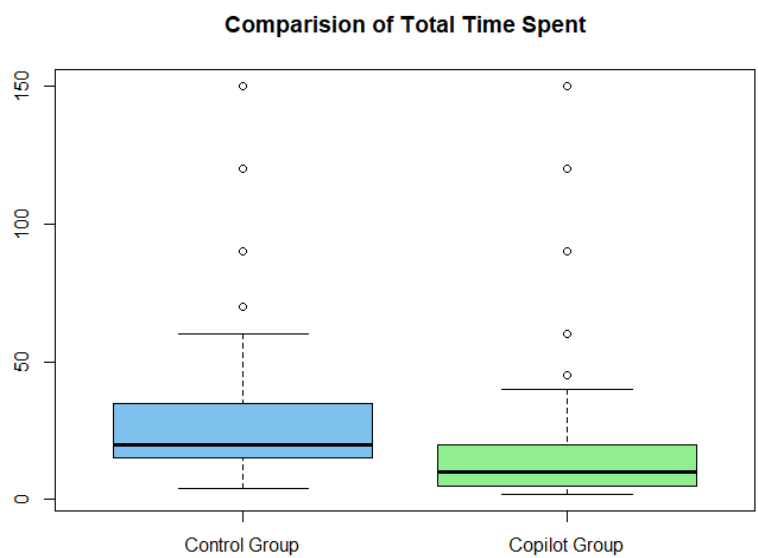

Figure 1. Boxplot for total time to solve a problem (Total_Time_Spent)

Figure 1 shows that the median value for time to solve the problem for Copilot Group is less than Control Group. Also, the range of valid Copilot data is much smaller than Control Group. According to this diagram, there are 9 outliers. However, in the analysis these data points were not excluded from the dataset.

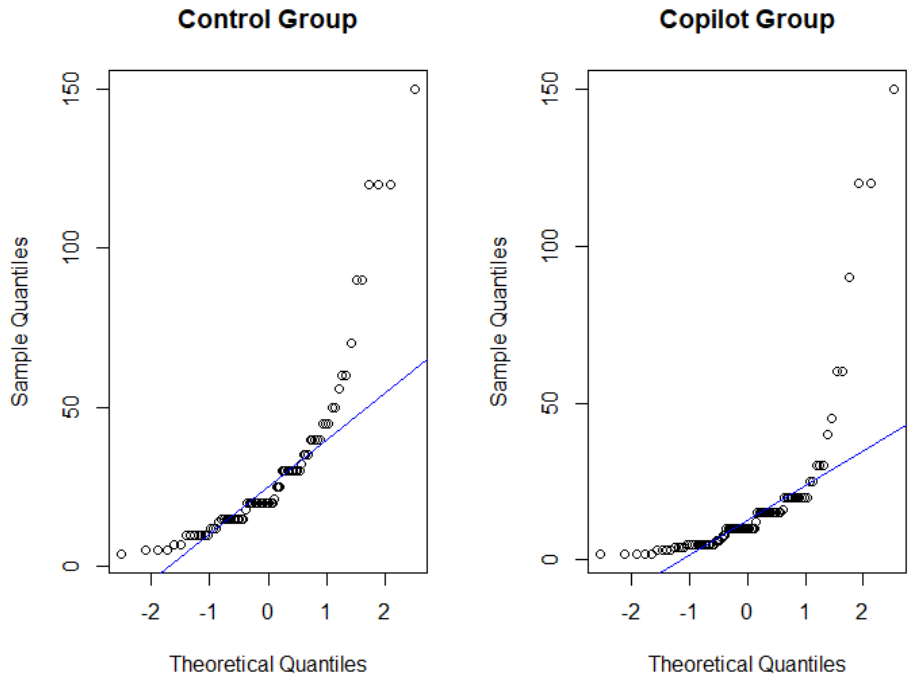

Figure 2. QQ-Plot for total time to solve a problem (Total_Time-Spent)

From the Quantile-Quantile (QQ) plot in Figure 2, it is evident that for both Control Group and Copilot Group, Total Time Spent does not follow normal distribution. Since the data does not follow normal distribution, data is skewed, and the number of data points is insufficient, non-parametric tests were performed on the data to test the A/B Testing hypothesis.

### 4.8. Non-parametric statistics

For non-parametric hypothesis test, Mann-Whitney U-test and Wilcoxon Signed Rank Test were considered. Finally, Wilcoxon Signed Rank Test were selected because Mann-Whitney U-test tests two independent samples, whereas the Wilcox sign test tests two dependent samples. The Wilcoxon Sign test is a test of dependency. In the case of Copilot experiment, Copilot Group and Control group consists same set of users. Hence the data related to Control group and Copilot group are not completely independent.

**Wilcoxon Signed Rank Test**
Wilcoxon Signed Rank Test were performed on data to test whether there is any significant difference between the Control Group and Copilot Group participants in terms of average time spent to solve a problem.

Below are the results of Wilcoxon Signed Rank Test results for the metrics selected as part of A/B Testing. The tests were one sided and Alpha value selected for this experiment is 0.05 (Confidence Interval 95%).

Table 2. Wilcoxon Signed Rank Test results

| Category | Metric | Hypothesis | Sample size | W | p-value | Decision |
|---|---|---|---|---|---|---|
| Productivity | Total_Time_Spent | *H0: As a result of using Copilot usage, there is no significant difference in terms of Total_Time_Spent for Copilot Group and Control Group*<br><br>*H1: As a result of using Copilot, Total_Time_Spent is less for Copilot Group* | 22 | 26 | 0.001 | H0 rejected |
| Code Quality | Unit_Test_Success_Ratio | *H0: As a result of using Copilot usage, there is no significant difference in terms of Unit_Test_Success_Ratio for Copilot Group and Control Group*<br><br>*H1: As a result of using Copilot, Unit_Test_Success_Ratio is higher for Copilot Group compared to Control Group* | 17 | 98 | 0.060 | H1 rejected |
| Code Quality | Bugs | *H0: As a result of using Copilot usage, there is no significant difference in terms of Bugs for Copilot Group and Control Group*<br><br>*H1: As a result of using Copilot, number of Bugs is less for Copilot Group compared to Control Group* | 17 | 0 | 0.033 | H0 rejected |
| Code Quality | Code_Smells | *H0: As a result of using Copilot usage, there is no significant difference in terms of Code_Smells for Copilot Group and Control Group* | 17 | 10 | 0.007 | H0 rejected |

| Category | Metric | Hypothesis | Sample size | W | p-value | Decision |
|---|---|---|---|---|---|---|
| | | H1: As a result of using Copilot, number of Code_Smells is less for Copilot Group compared to Control Group | | | | |
| Code Security | Vulnerabilities | H0: As a result of using Copilot usage, there is no significant difference in terms of Vulnerabilities for Copilot Group and Control Group | The SonarQube code review generated only 1 non-zero data point against Vulnerabilities. There is not enough data available to test this metric. | | | |
| | | H1: As a result of using Copilot, number of Vulnerabilities is less for Copilot Group compared to Control Group | | | | |

Based on the data collected as part of the A/B Testing, results of Wilcoxon Signed Rank Test in Table 2 suggests that, because of using Copilot there are significant improvements on productivity and code quality related metrics. However, The SonarQube code review did not produce sufficient evidence to make any conclusion on Security aspect of the code submitted by the participants.

### 4.9. Results

#### 4.9.1. Productivity

Table 3: Productivity increase across different proficiency levels

| Python proficiency | Mean Total_Time_Spent (Control Group)/per problem (in minutes) | Mean Total_Time_Spent(Copilot Group)/per problem (in minutes) | Productivity Improvement |
|---|---|---|---|
| Beginner | 20.07 | 9.58 | 52.27% |
| Intermediate | 28.60 | 16.70 | 41.6% |
| Advanced | 39.82 | 23.70 | 40.48% |

Table 3 describes a closer look how the productivity increased among participants with different Python Proficiency levels between Control group and Copilot Group. Since there are only few participants in "Novice" and "Expert" Python proficiency levels, the participants with "Novice" proficiency have also been considered as "Beginner" proficiency, similarly "Expert" proficiency level participants have also been considered as "Advanced" in the table below.

From Table 1,

> ***Overall Productivity Improvement =***
> ***((Mean Total_Time_Spent(Control Group) - Mean Total_Time_Spent(Copilot Group)) / Mean Total_Time_Spent(Control Group) )*100= ((30.98-17.86)/30.98)*100 = 42.34%***

This study shows that Copilot improves the productivity of ANZ engineers by 42.36% on an average. As can be seen in Figure 3, Copilot users took less time overall on each challenge problem than participants in the 'Control group'.

Assessment of productivity based on Python proficiency found Copilot was beneficial to participants for all skill levels but was most helpful for those who were 'Expert' python programmers, Figure 3. For all code challenge levels from 'Very Easy' to 'Hard', the Copilot group spent less time on average than Control group completing the task, Figure 4. As expected, Copilot use gave the largest improvement when completing 'Hard' tasks.

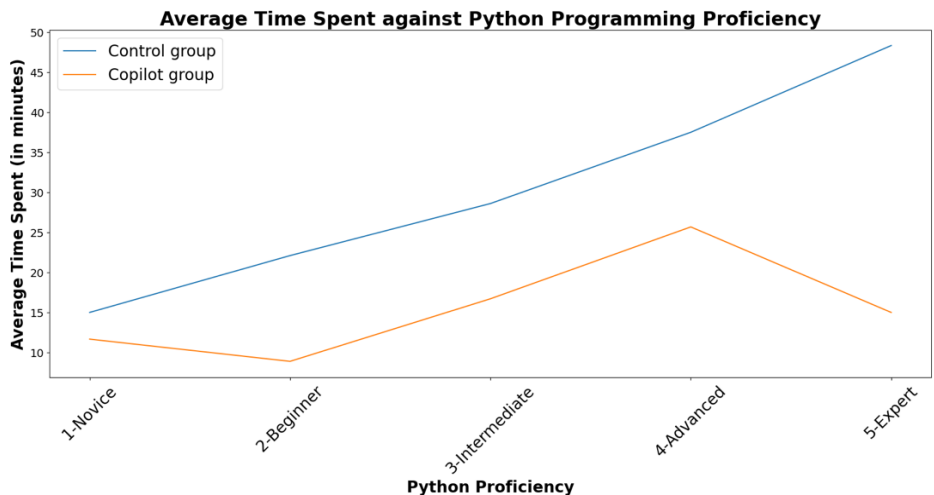
Figure 3. Average Time Spent to Solve Problems vs Python Proficiency

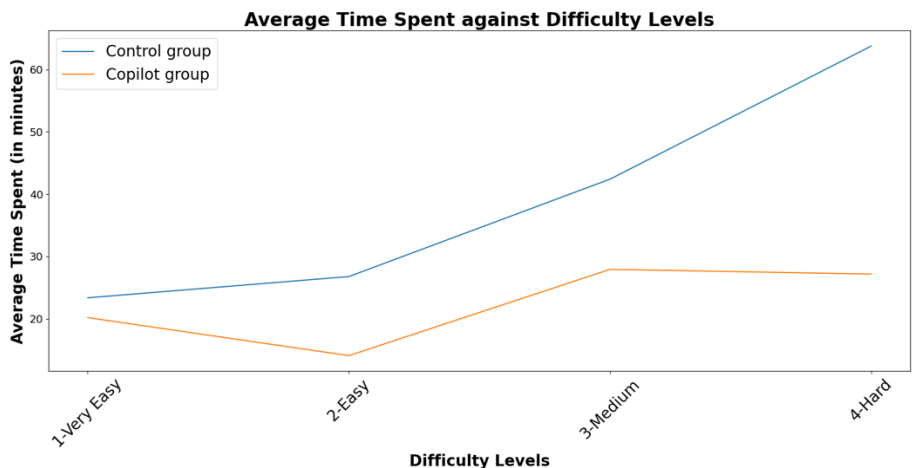
Figure 4. Average Time Spent to Solve Problems vs Problem Difficulty Levels

**4.9.2. Quality**

According to the unit test results, methods written by Copilot Group participants had 12.86% more success ratio than methods written by Control group users. However, this result is not statistically significant. Table 3 suggests that participants with "Beginner" Python proficiency received the highest benefit from using Copilot.

### 4.9.3. Security

This section covers how engineers - with and without Copilot - performed from a security standpoint, providing insight into both the introduction of security risks as well as minimisation of existing ones. This aspect was assessed through the presence of vulnerabilities in the code submitted by participants. To test secure coding practices, one security-related question was inserted per week of the A/B testing phase:

"password usecase" prompted users to create a function that would securely hash an inputted password using the pbkdf2_hmac function from the python 'hashlib' module. Key to this question was the use of a randomly generated cryptographic salt which would be passed into the hash function. The randomness of this salt was stipulated in the problem instructions; in the real world, if salts aren't randomly generated, hackers are better able to match the inputs and outputs of a hash function. The Sonar Way contains a vulnerability definition for non-random cryptographic salts, allowing participants' solutions to be checked for this.

"code executor usecase" prompted users to create a basic FastAPI application that would accept a 'command' parameter within POST requests to its '/execute' endpoint. These commands would correspond in name to functions that the application would need to run. The purpose of the question was to ensure participants ran checks on the command parameters, prior to running the functions. Otherwise, their code would be susceptible to injection attacks. The Sonar Way contains a vulnerability definition for code that is susceptible to code injection attacks, allowing this to be assessed.

Vulnerabilities were scanned for in participants' solutions to questions they received in the A/B testing phase, using SonarQube's static code analysis. SonarQube defines a vulnerability as "a security-related issue that represents a backdoor for attackers."

### 4.9.4. Sentiment Around Copilot

The following sections evaluate how engineers' experiences with GitHub Copilot varied according to different independent variables. They offer an idea of how different levels of programming, Visual Studio Code, and GitHub Copilot experience correlate to different qualities of user experience with the tool; and they offer insight into what tasks GitHub Copilot may lend itself most effectively to.

All values in the tables are median values, i.e. the 50th percentile response from participants. The table cells are colour-coded according to their favourability towards Github Copilot:

Firstly, shown below is an aggregation of all survey responses received over the course of phase 1 (weeks 1-2). Across all areas, participants responded positively regarding GitHub Copilot. They felt it helped them review and understand existing code, create documentation, and test their code; they felt it allowed them to spend less time debugging their code and reduced their overall development time; and they felt the suggestions it provided were somewhat helpful, and aligned well with their project's coding standards. It should be noted, however, that the magnitude of each area's sentiment fell short of the "strong positive"

Table 4: Participant sentiment across surveyed areas of code development

| Area | Aggregate Perceived Effect |
|---|---|
| Time spent debugging code using Copilot | A bit less time |
| Time needed to produce the same code without Copilot | A bit more time |
| Suggestions' alignment with project's coding standards | Well |
| Quality of suggestions received | Somewhat helpful |
| Impact on ability to review and understand existing code | Positive effect |
| Impact on ability to create unit tests for code | Positive effect |
| Impact on ability to create documentation of the code | Positive effect |

## 5. DISCUSSIONS

### 5.1. Limitations

**Sample size**

A notable limitation of our experiment was the level of engagement, which impacted the robustness of the conclusions drawn. While over 100 engineers participated over a six-week period, engaging in a variety of tasks, participation rates fluctuated significantly across these tasks. This variability in engagement levels presents a challenge in generalizing the findings.

Moreover, considering that ANZ Bank employs around 5000 engineers in diverse roles, the sample size of our experiment represents only a fraction of this population. Therefore, our findings, while indicative, may not fully encapsulate the broader impact and potential of GitHub Copilot across the entire spectrum of the organization's software engineering workforce.

**Programming questions**

Some programming questions, such as "password usecase" were considered by some participants to be difficult to understand/implement, and consequently received fewer attempts. A lower sample size means less reliable results when comparing participants' attempts at these questions.

There must also be a discussion of the SonarQube scan results and their relation to the programming questions. As mentioned in the "Key Constraints" section, the programming questions were designed to be light weight so that participants could complete them alongside their daily work. By the same token, however, the relatively small amount of code required to solve the questions did not lend itself to a critical static code analysis; there was little room for bugs and vulnerabilities to exist, as the questions were atomic and short. While some data was obtained regarding code smells in the code belonging to both groups, the other two categories could not be reasonably measured by the code scan due to the nature of the tasks that the participants had been given. If this experiment were to be repeated, a project-style task, reflecting a more concrete development goal, might be more suitable; this would prompt participants to write a larger volume of code to tackle a single problem that would more closely resemble a workplace task.

**Biases**

Biases such as the Dunning-Kruger effect may also affect the accuracy of self-reported proficiency levels in programming languages throughout the experiment. It is possible that programmers of low-proficiency overestimated their ability in their chosen language or python, while high-proficiency programmers underestimated their ability. With this said, the bias can be safely assumed to exist to an equal extent in both the Copilot and control groups (as they were created to have the same composition using a random selection process within each experience level), meaning the relative comparisons between experience levels remain valid.

Similarly, participants may have under-reported performance metrics such as the time spent developing and debugging their code – the effect of this can is also negated by the assumption that this tendency was equally present in both groups.

Biases may have affected sentiment data surrounding Copilot, as it is likely that those who volunteered for the Copilot experiment did so because they were interested or enthusiastic about the tool. It is also possible that participants joined to demonstrate that the tool did not influence their performance, but it seems likely that the positive bias would outweigh this negative bias. It is therefore possible that the positive overall sentiment results may be a consequence of the participant population's positive predisposition toward the tool.

## 5.2. Summary and Future Work

This paper presents evidence on the impacts that GitHub Copilot may have on productivity, code quality and code security in ANZ Engineering. The results suggests that Copilot has a statistically and practically significant impact on productivity and code quality. The group that had access to GitHub Copilot was able to complete their tasks 42.36% faster than the control group participants. This result is statistically significant. The code produced by Copilot participants contained fewer code smells and bugs on average, meaning it would be more maintainable and less likely to break in production. These observations were also shown to be statistically significant. The experiment could not generate meaningful data which would measure code security. However, the data suggest that Copilot did not introduce any major security issues into the code.

The experiment yielded conclusive results regarding user sentiment toward GitHub Copilot. Participants felt it had a "positive effect" on their ability to review and understand existing code, create documentation of their code, and create unit tests for their code. They felt it helped them spend "a bit less time" debugging their code and that they would have spent "a bit more time" producing the same code without GitHub Copilot. They felt the suggestions they received were "somewhat helpful" and aligned with their coding standards "well". While the sentiments were uniformly positive in valence, they were all moderate; in none of the surveyed areas did the median participant respond with the maximum degree of positivity. The qualitative feedback from the participants suggests that there are areas of improvement for Copilot to be more effective in improving the developer's experience.

However, considering the quantitative and qualitative analysis of the data generated in this experiment and subject to further analysis on security of the code suggested by Copilot, it is recommended to productionise GitHub Copilot at ANZ Bank.

In conclusion, this research provides compelling evidence of the transformative impact of GitHub Copilot on engineering practices at ANZ Bank. The adoption of this tool has marked a shift in the work paradigm, empowering engineers to focus more on creative and design tasks while reducing time spent on repetitive coding. As of the writing of this paper, GitHub Copilot has already seen significant adoption within the organization, with over 1,000 users using it into their workflows. Concurrently, a detailed investigation into the productivity improvements attributable to GitHub Copilot is underway. This ongoing study aims to quantify the tool's impact on operational efficiency and overall performance at ANZ Bank.

## AUTHORS

**Name: S**ayan Chatterjee
As a Data Architect at ANZ Bank, he possesses a distinguished academic profile, having earned a Master of Data Science from the University of Melbourne and a Bachelor of Engineering from Indian Institute of Engineering, Science & Technology, Shibpur. His professional journey extends beyond conventional academia, having actively contributed to research projects at the Indian Statistical Institute (ISI) and Variable Energy Cyclotron Centre (VECC). Specifically, he is adept at navigating the application of Large Language Models (LLMs) in the corporate environment, demonstrating a nuanced understanding of how these cutting-edge technologies can be harnessed to enhance data architecture and decision-making processes.

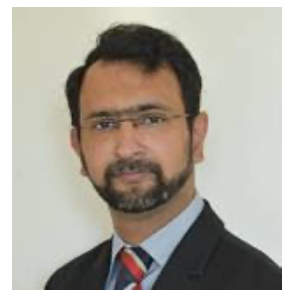

**Name:** Ching Louis Liu

Dr. Ching Louis Liu earned his Ph.D. in Software Engineering with a focus on Artificial Intelligence from The University of Melbourne. He currently serves as the Senior Manager of AI and Data Analytics in the Engineering division at ANZ Bank. Beyond his role in the IT industry, Dr. Liu is deeply engaged in academic pursuits and education. He contributes to the academic community by teaching and developing curriculum for various postgraduate degrees in Data Science and Software Engineering at RMIT University. Additionally, Dr. Liu supervises Master's students at the University of Bath in the UK, further demonstrating his commitment to fostering the next generation of professionals in his field

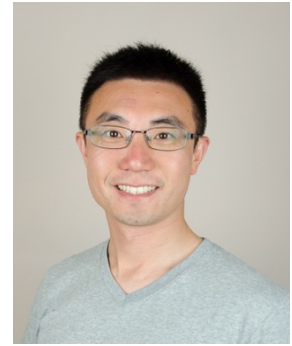

**Name:** Gareth Rowland

Gareth has worked in the Technology industry for almost 40 years originally developing embedded manufacturing equipment controllers and enterprise resource planning solutions. He subsequently qualified as a Chartered Accountant in the UK and combined his technical knowledge with a strong level of business acumen to provide consulting services through KPMG for almost 15 years. Over the past ten years Gareth has worked in enterprises to transform their Continuous Delivery capabilities and building collaborative environments for Engineering teams. and establish measures that can help these teams. Gareth has an active interest in Data Science (including Analytics, Machine Learning and Gen-AI) having recently completed his Masters in Data Science Strategy and Leadership (RMIT). He is applying this study to research how Software Engineering is likely to change in the future as a result of these new technologies.

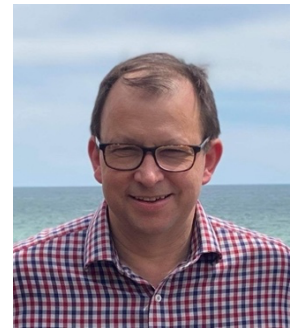

**Name:** Tim Hogarth

Tim Hogarth is the Chief Technology Officer at ANZ and is accountable for the architecture of the bank, the engineering standards and plays a key role in the technology planning and prioritisation.  Tim works closely with business teams, helping shape roadmaps and aligning technology investment.  He also works closely with the architecture and engineering teams at ANZ, driving up the bank's technology prowess through a focus on pragmatic standards, common platforms and engineering excellence.

Tim's big drivers and interests are anchored to AI, cloud migration, better management of data and using sophisticated engineering to drive up resilience and reduce complexity.

He's worked in other banks in North America, the UK and in Australia, and has a technology delivery track record spanning more than twenty years. A practical and considered leader, Tim spends a great deal of time thinking about how to harness emerging technology industry shifts for greater enterprise value in large organisations.

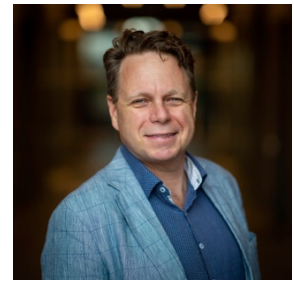